\documentclass[preprint]{elsarticle}
\usepackage{lineno}
\usepackage{adjustbox}
\usepackage{xcolor}
\usepackage{bm}
\usepackage{subfig}
\usepackage{mathtools}
\usepackage{amsmath}
\usepackage{graphicx}
\usepackage{array}
\newcommand{\Up}{\mid \Uparrow \rangle}
\newcommand{\Down}{\mid \Downarrow \rangle}
\newcommand{\up}{\mid \uparrow \rangle}
\newcommand{\down}{\mid \downarrow \rangle}

\newcommand{\BETA}{|\beta\rangle_{A,{\cal E}}}
\newcommand{\PLRZ}{\left|\frac{N_A}{2},0\right\rangle_A}
\newcommand{\R}{|R\rangle_A}
\begin{document}

\begin{frontmatter}
\title{Quantum dynamics of a small symmetry breaking measurement device}
\author[nij]{H.~C.~Donker}
\author[gro]{H.~De Raedt}
\author[nij]{M.~I.~Katsnelson\corref{cor1}}
\ead{M.Katsnelson@science.ru.nl}
\cortext[cor1]{Corresponding author}
\address[nij]{Radboud University, Institute for Molecules and Materials, Heyendaalseweg 135, NL-6525AJ Nijmegen, The Netherlands}
\address[gro]{Zernike Institute for Advanced Materials, University of Groningen, Nijenborgh 4, NL-9747 AG Groningen, The Netherlands}

	\begin{abstract}
	A quantum measuring instrument is constructed that utilises symmetry breaking to enhance a microscopic signal. The entire quantum system consists of a system-apparatus-environment triad that is composed of a small set of spin-1/2 particles. The apparatus is a ferromagnet that measures the $z$-component of a single spin. A full quantum many-body calculation allows for a careful examination of the loss of phase coherence, the formation and amplification of system-apparatus correlations, the irreversibility of registration, the fault tolerance, and the bias of the device.
	\end{abstract}
\end{frontmatter}

%\linenumbers

\section{Introduction}
Bohr, in his discussions with Einstein~\cite{BOHR49}, repeatedly emphasised that each peculiar feature or seemingly paradoxical phenomenon in quantum mechanics, is always to be viewed in the light of the experimental arrangement that is used to interrogate the quantal test object (e.g., an electron or atom). 
But if one takes seriously the idea that the laws of quantum mechanics are universally valid, then, in turn, the measurement instrument itself must obey the laws of quantum mechanics.
The first attempt to consolidate the internal consistency of measurement and quantum theory was by von Neumann~\cite{NEUM55}. 
By detailing measurement as a non-unitary disruption of the density operator to diagonal form (in the basis determined by the measurement), von Neumann noted that the same result can be accomplished by considering an enlarged quantum system in which the Hilbert space is partitioned in two ${\cal H}_{AB} = {\cal H}_A\otimes {\cal H}_B$~\cite{NEUM55}. Particular interactions between $A$ and $B$ entangle initially uncorrelated quantum states, and the density operator pertaining to $A$ leads to diagonal form after coarse graining over $B$, the basis of which is determined by the form of the interaction.
These foundational works have led to new flourishing fields of research---notably the theory of decoherence~\cite{JOOS03,ZURE03, SCHL07}---and by now the quantum theory of measurement comprises a vast amount of work, see e.g. Refs.~\cite{WHEE83,ALLA13}.

One of the salient features of quantum measurement is the ability to amplify signals. This was stressed in, e.g., Refs.~\cite{vKAMP88,ZIMA88,GAVE90, ALLA03, ALLA13} and amplification is now also adopted by decoherence theory under the umbrella of quantum Darwinism~\cite{ZURE09}. One particular way to achieve amplification of a quantum signal is to utilise the sensitivity of a system that is prone to symmetry breaking~\cite{ZIMA88,GAVE90,FIOR94,ALLA03,LAUG06}. One frequently encountered example that is used to illustrate~\cite{GREE58,ZIMA88,GAVE90,FIOR94,GRAD94} measurement as a phase transition is Wilson's cloud chamber. 
This device contains a metastable gas where droplets are formed upon ionisation of atoms, which lead to particle tracks. 
In analogy, the same rational of phase nucleation in a metastable state can be used to enhance microscopic perturbations in a variety of other host materials. Particular focus has been on magnetic systems~\cite{GAVE90, FIOR94, MAYB95, ALLA03, ALLA13}.
The requirement that the detector is metastable, which sowed the seeds of a symmetry breaking instrument, can already be traced back to~\cite{GREE58,BLOK68}. 
Although several other works discuss phase transitions and symmetry breaking in some relation to quantum measurement~\cite{MERM91,GRAD94,ANDE97,MERL06,LAND13}, the most relevant works for this paper are Refs.~\cite{GAVE90,ALLA03} and in particular Ref.~\cite{ALLA13} where a magnetic set up was used to examine the dynamics of a symmetry breaking measurement device. For a more comprehensive overview of quantum measurement models the reader is referred to~\cite{ALLA13}.

This work aims to complement the detailed work of Refs.~\cite{ALLA03,ALLA13} by studying the full quantum many-body dynamics (instead of a mean field model) of a ferromagnetic apparatus undergoing symmetry breaking.
To this end, the time-evolution of a few-particle apparatus is considered that measures a spin 1/2. The apparatus consists of a ferromagnet that is in contact with a thermal reservoir. The goal is to examine whether a fully quantum mechanical apparatus of modest size is indeed able to capture the most important measurement aspects, such as truncation (i.e., loss of phase coherence) of the test object, correlation with the apparatus, amplification of the signal, and reliable registration of the outcome.

This paper is organised as follows: First, the Hamiltonian of our device is laid out in Sec.~\ref{sec:H}. The initial states and the numerical implementation of the model are discussed in Sec.~\ref{sec:sim}. 
Next, the correlation development with the test object and the decoherence of the apparatus are analysed in Sec.~\ref{sec:truncation}.
Then, in Sec.~\ref{sec:calibration}, the bias and predictability of the device is examined. The stability and the irreversibility of the instrument is further investigated in Sec.~\ref{sec:stability}. And finally, some recapitulating and concluding marks are made in Sec.~\ref{sec:disc_concl}.

\section{Method}
\begin{figure}
	\centering
	\includegraphics[scale=0.2]{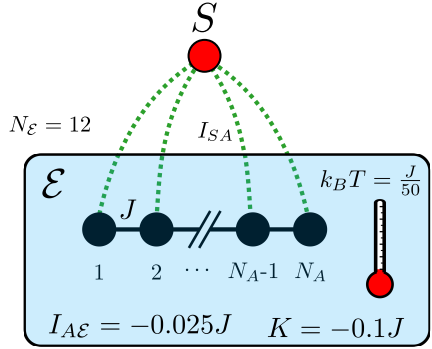}
	\caption{Schematic of the measurement set up. A ferromagnet of $N_A$ spins 1/2 with exchange constant $J$ is immersed in a spin-glass like environment ${\cal E}$ of $N_{\cal E}=12$ spins that resemble a thermal reservoir of temperature $\beta^{-1}=J/50$. The intra-environment strength is $K$ and the environment interaction with the ferromagnet is of size $I_{A{\cal E}}$. The order parameter of the ferromagnet is coupled to $S$ along the $z$-direction with strength $I_{SA}$ so as to measure its spin.}
	\label{fig:setup}
\end{figure}
\subsection{Hamiltonian}\label{sec:H}
The entire system is composed of a collection of spin-1/2 particles, in which a measurement instrument measures spin $S$ in the von Neumann sense~\cite{NEUM55}. The apparatus is composed of a $N_A$ ferromagnetically coupled spins, collectively called $A$, that are in contact with an environment ${\cal E}$, of $N_{\cal E}$ spins. For simplicity the self-Hamiltonian of $S$ is neglected.
The entire Hamiltonian is

\begin{equation}\label{eq:H}
H = H_{SA} + H_A + H_{A{\cal E}} + H_{\cal E} \, ,
\end{equation}
with $H_{SA}$ ($H_{A{\cal E}}$) the system-apparatus (apparatus-environment) interaction and $H_A$ ($H_{\cal E}$) the self-Hamiltonian of the apparatus (environment). The Hamiltonian is schematically depicted in Fig.~\ref{fig:setup}.
The apparatus consists of an open Heisenberg ferromagnetic chain~\cite{MAHA00}
\begin{equation}\label{eq:HH}
H_A = -J \sum_{i=2}^{N_A} \bm{S}_i \cdot \bm{S}_{i+1} \, ,
\end{equation}
in which the spin indices are labelled from $i=2 \dots N_A + 1$, $S_l^\alpha = \sigma_l^\alpha/2$ are spin-1/2 operators in terms of Pauli matrices, and $J > 0$ to ensure ferromagnetism. The basic premise of the measurement device is that the initial unstable configuration, the ready state, is sensitive to perturbations of the order parameter $\bm{S}_A = \sum_{i=2}^{N_A+1} \bm{S}_i$. The device will be used to measure the $z$-direction of $S$, therefore the coupling
\begin{equation}\label{eq:H_SA}
H_{SA} = -I_{SA} S^z_S S_A^z \, ,
\end{equation}
of strength $I_{SA}$ is used, which ensures that $z$-direction measurements are non-destructive in that direction. The minus sign in $H_{SA}$ will lead to parallel alignment, and therefore positive correlation along the $z$-direction, between the system and the apparatus when $I_{SA} > 0$.
For the environment, a spin glass-like environment
\begin{equation}
H_{\cal E} = K \sum_{\alpha \in \{x,y,z\}}\sum_{k,l \in {\cal E}}  r_{kl}^{\alpha} S^\alpha_k S^\alpha_l \, , 
\end{equation}
is used to facilitate decoherence and relaxation in $A$~\cite{YUAN07,YUAN06}, with $K$ the interaction strength, $r_{kl}^{\alpha}$ are random numbers uniformly distributed in the range $[-1, 1]$ and the sum is over all spins in ${\cal E}$.
The apparatus-environment coupling consists of 
\begin{equation}
H_{A{\cal E}} = -I_{A{\cal E}} \sum_{i \in A,k \in {\cal E}} r_{ik} \bm{S}_i \cdot \bm{S}_k \, ,
\end{equation}
with $r_{ik}$ uniform in $[0,1]$ and the summation $i$ ($k$) over all spin indices in $A$ (${\cal E}$).
The importance of an environment besides the apparatus is multifold, e.g., to remove the system-apparatus basis ambiguity~\cite{ZURE81, ZURE03, SCHL07} and to enhance irreversible registration~\cite{ALLA13}. Here, in addition, it is one of practical interest: in order to get pointer readings, terms not commuting with the order parameter are needed in the Hamiltonian.

\subsection{Initial states and simulation procedure}\label{sec:sim}
As indicated in the Introduction, one of the primary goals of a measurement instrument is to become correlated with the test object. It will therefore be assumed that initially, there is no correlation between the apparatus $A$ and the test spin $S$. 
This is ensured by writing the wave function at $t=0$ as a product state
\begin{equation}
|\Psi(t=0)\rangle = |\psi_0\rangle_S \otimes |0\rangle_{A,{\cal E}} \, ,
\end{equation}
with $|\psi_0\rangle$ the initial state of $S$ (indicated by the subscript) and $|0\rangle_{A,{\cal E}}$ the \emph{ready-state} of the apparatus and environment combined. More specifically, for $|\psi_0\rangle$ the following family of initial states of $S$
\begin{equation}\label{eq:S_a}
|\psi(a)\rangle_S = \sqrt{a}\up + \sqrt{1-a}\down \, ,
\end{equation}
parametrised by $a \in [0,1]$ will be examined [additional phase factors are unimportant in Eq.~(\ref{eq:H_SA})]. Furthermore, three different ready states $|0\rangle_{A,{\cal E}}$ are analysed. First off, a product state $\PLRZ \otimes |\beta\rangle_{\cal E}$ with 
\begin{equation}
\PLRZ = \left(S_A^-\right)^{N_A/2}\Up_{A}  \, ,
\end{equation}
the ferromagnetic state of maximal multiplicity $S_A=N_A/2$ in which $S_A^z\PLRZ=0$  ($\Up_{A}$ is the fully polarised state with all spins in $A$ up) and where the environment is in a thermal-like state $|\beta\rangle_{\cal E}$ of inverse temperature $\beta$~\cite{HAMS00}. The latter is constructed by first generating Gaussian-distributed random weights for each element of the wave function in ${\cal E}$, using the Box-Muller method~\cite{BOX58}. 
Next, imaginary time-evolution $\exp[-\beta H_{\cal E}/2]$ is carried out on the normalised random state to project onto a low energy configuration~\cite{HAMS00}. And finally, the resulting wave function is normalised to give $|\beta\rangle_{\cal E}$. Observe that $\PLRZ$ belongs to the $N_A+1$ fold degenerate ground state subspace of $H_A$ [Eq.~(\ref{eq:HH})]. In an isolated ferromagnet, the state $\PLRZ$ can therefore be carried to the fully polarised states $\Up_{A}$ and $\Down_{A}$ without energy cost.

Secondly, the ready state $|R\rangle_A \otimes |\beta\rangle_{\cal E}$ will be examined in which the states in the $S_A^z=0$ subspace ($d^0_A=N_A!/(N_A/2)!^2$ in total) have Gaussian random weight in $|R\rangle_A$. The environment state $|\beta\rangle_{\cal E}$ is the same as before. The aim of this state is to examine impact of the the energy content of the state, since $|R\rangle_A$ resembles an infinite temperature ($\beta=0$) state in the subspace spanned by vanishing order parameter states.

And thirdly, the combined state $\BETA$ that is like $|\beta\rangle_{\cal E}$ but now for both $A$ and ${\cal E}$. That is, it is constructed from a Box-Muller state before projecting with $\exp[-\beta(H_A+H_{A{\cal E}}+H_{\cal E})/2]$. From a practical point of view this state is perhaps most realistic, in that the apparatus is initially in thermal equilibrium with its surrounding, rather than being isolated (as for the previous two uncorrelated configurations).

For convenience, the initial state $\PLRZ \otimes |\beta\rangle_{\cal E}$ ($|R\rangle_A \otimes |\beta\rangle_{\cal E}$) shall henceforth be abbreviated as $\PLRZ$ ($|R\rangle_A$); the suppressed $|\beta\rangle_{\cal E}$ is implicit. Of the three ready states, the order parameter $S_A^z$ of $\PLRZ$ and $\R$ are by construction unbiased. 

To calculate the time evolution of the wave function, the operator $\exp[-itH]$ is expanded in Chebyshev polynomials upto machine precision~\cite{DOBR03,RAED06}. Throughout this work units in which $\hbar=1$ and $k_B=1$ are used. 
The resulting quantities corresponding to a specific initial state are thus obtained via $|\Psi(t)\rangle = \exp[-itH]|\Psi(0)\rangle$.
The imaginary time operator is calculated similarly, with the addition of a subsequent wave function normalisation step.
The simulation data shown in this paper correspond to a single, but representative, realisation of the Hamiltonian couplings. 
The efficiency of the measurement device is quite sensitive to the precise numerical values of the coupling constants. As analysed in~\ref{sec:tuning}, the constraints on the coupling strengths is a consequence of the number of spins, which are expected to disappear in a sufficiently large system.
For convenience, the simulation parameters used throughout this work (and its respective values) are summarised in Tab.~\ref{tab:parameters}.

\section{Decoherence and development of correlations}\label{sec:truncation}
\begin{table}
	\centering
	\begin{tabular}{c|c|c|c|c}
	$\bm{I_{SA}}$ & $\bm{I_{A{\cal E}}}$ & $\bm{K}$ & $\bm{\beta}$ & $\bm{N_{\cal E}}$\\
	\hline
	$0.25 J$ & $-0.025 J$ & $-0.1 J$ & $50.0/J$ & 12\\
	\end{tabular}
	\caption{Parameters used in the simulations.}
	\label{tab:parameters}
\end{table}

\begin{figure*}
\begin{adjustbox}{center}
	\begin{tabular}{m{0.03\textwidth}|*{3}{m{0.38\textwidth}}}
		& \centering{$N_A=4$} & \centering{$N_A=6$} & \hspace{2cm}$N_A=8$ \\
		 \hline 
		 \renewcommand{\arraystretch}{0.0}% Tighter
		\rotatebox{90}{Correlation} &  %\vspace*{0.cm}
		\begin{center}
			\includegraphics[width=0.39\textwidth]{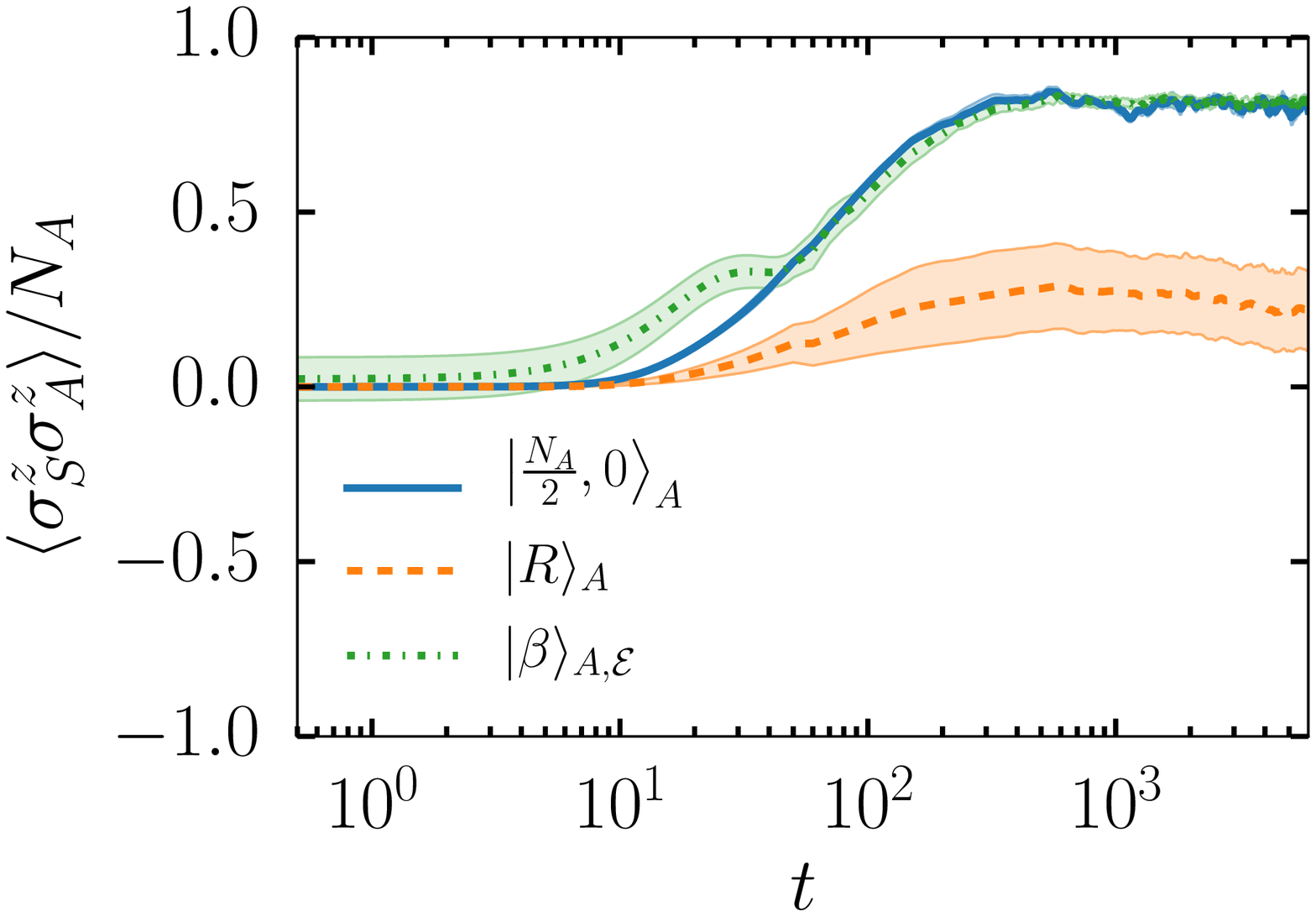}
		\end{center} 
		&
		\begin{center}
			\includegraphics[width=0.39\textwidth]{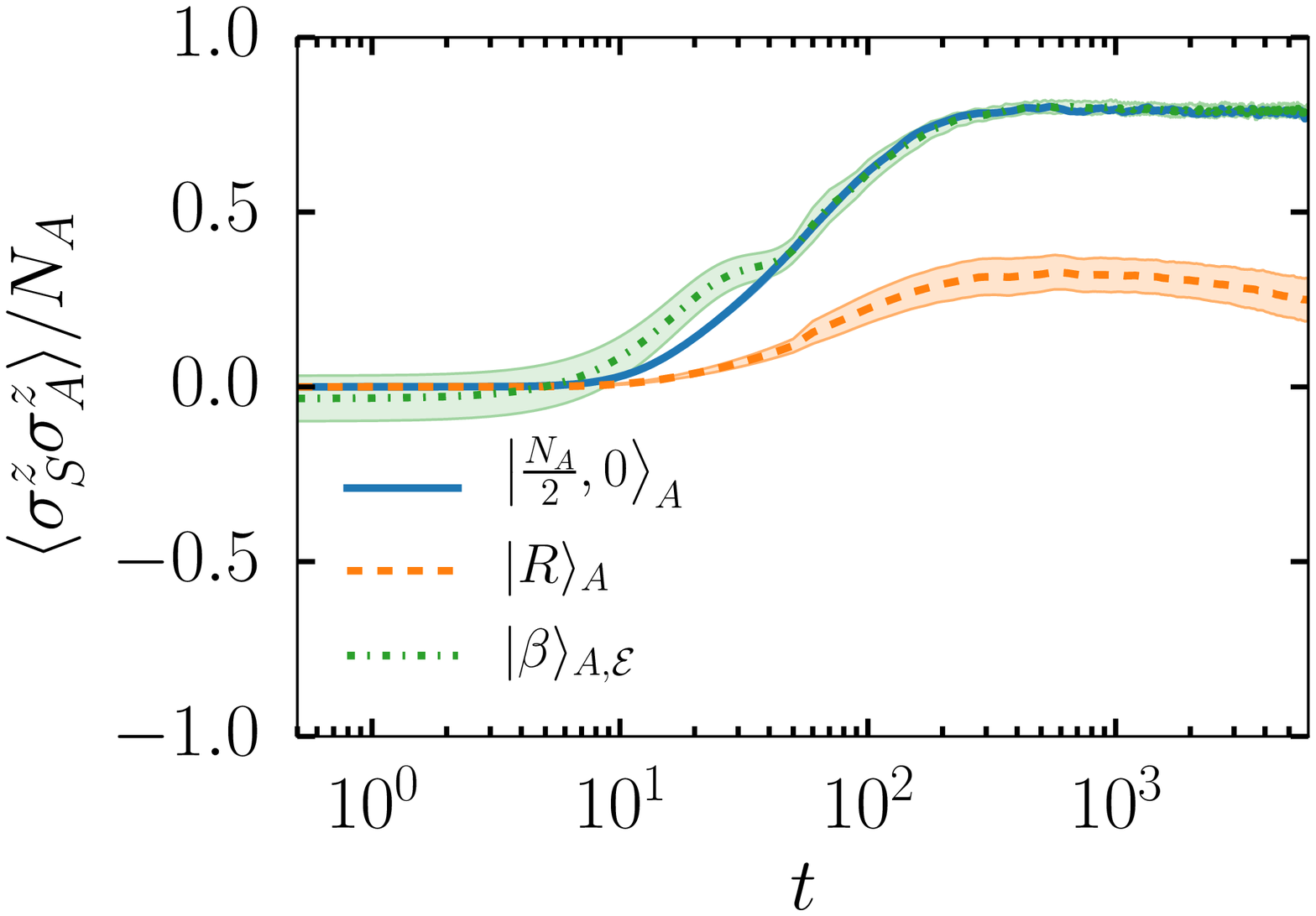}
		\end{center} 
		 &
		\begin{center}
			\includegraphics[width=0.39\textwidth]{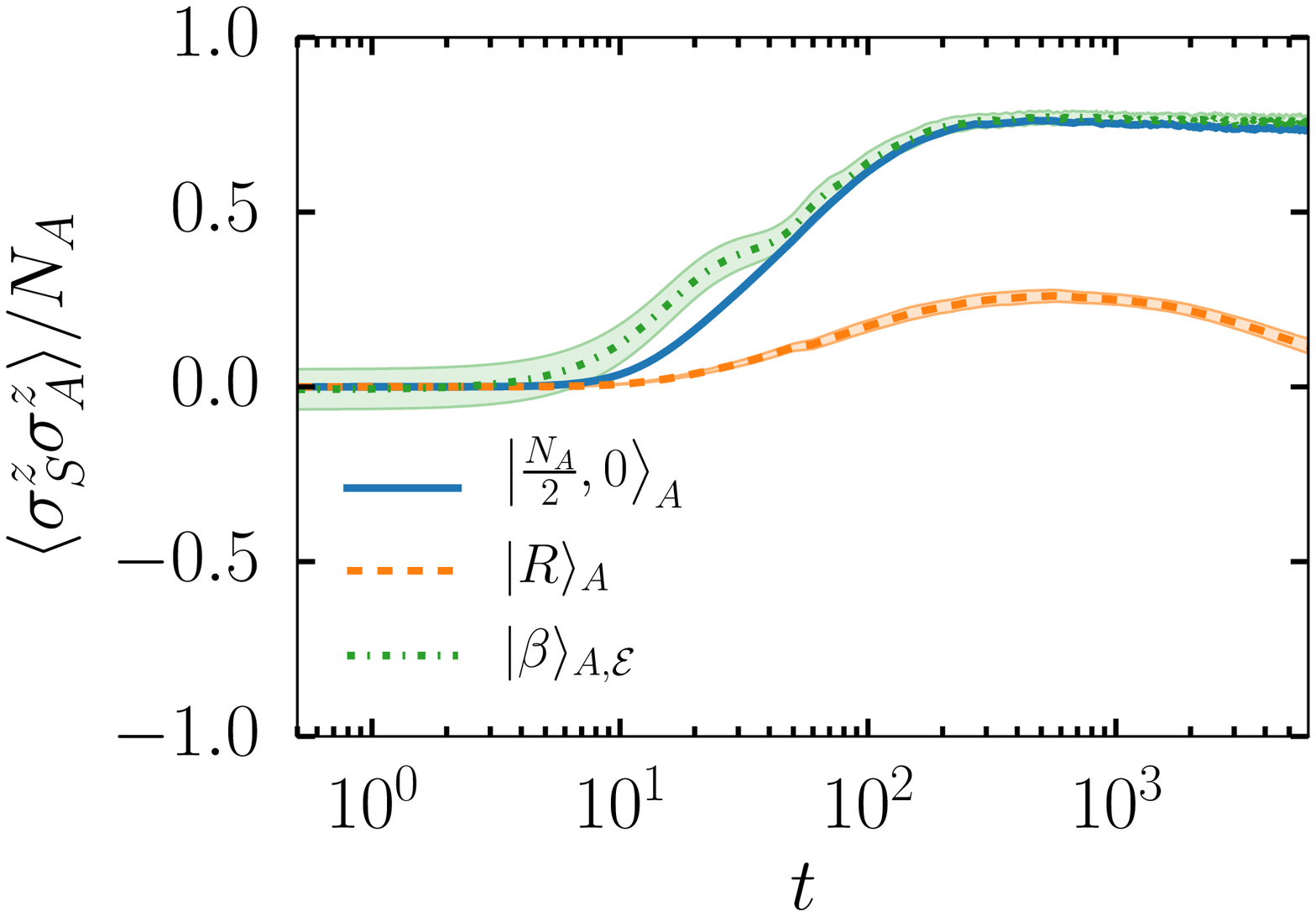}
		\end{center}
		\\
		%\hline
		\rotatebox{90}{Coherence} & 
		\vspace*{-1.3cm}
		\begin{center} 
			\includegraphics[width=0.39\textwidth]{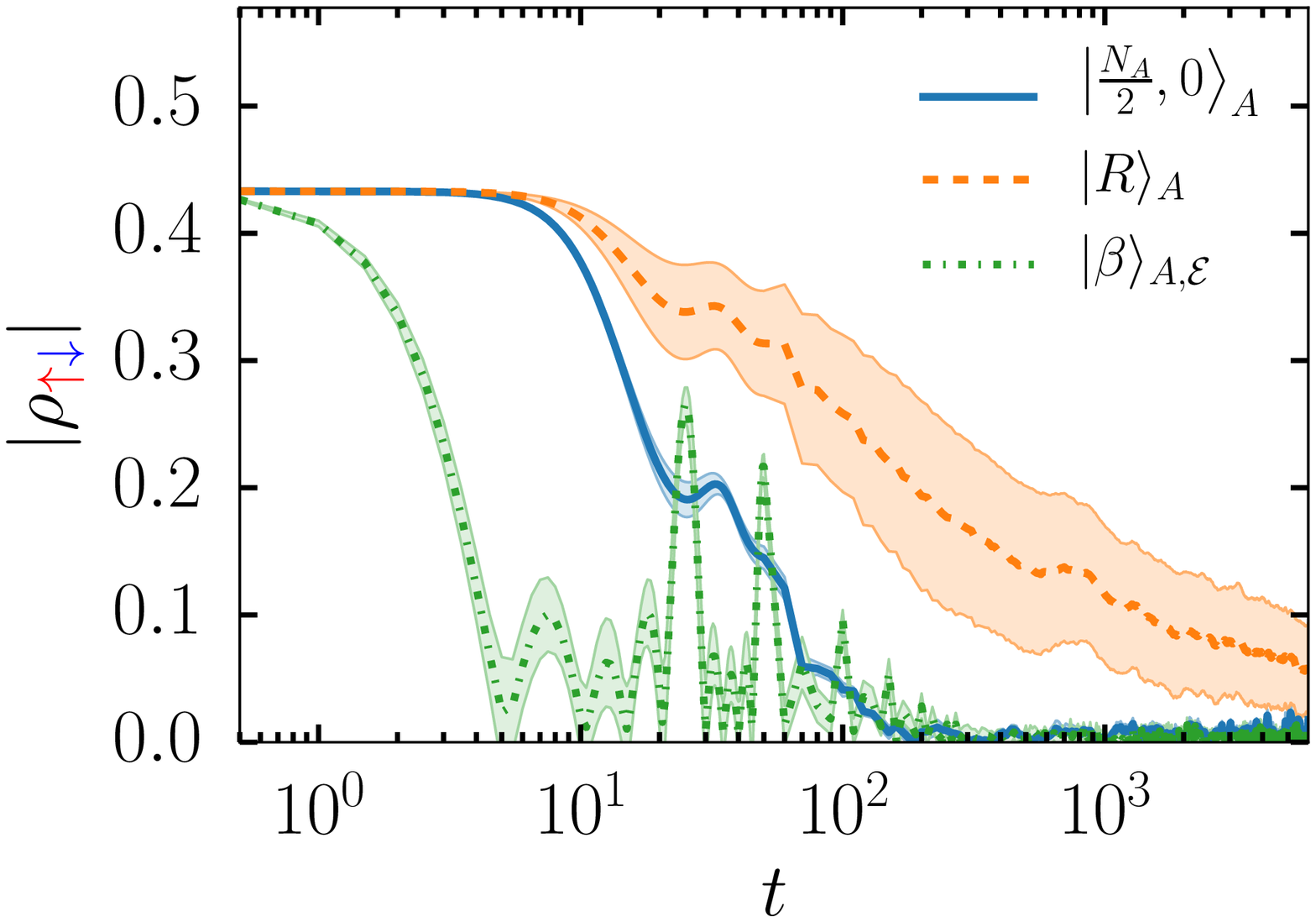}
		\end{center} 
		&
		 \vspace*{-1.3cm} 
		\begin{center}
			\includegraphics[width=0.39\textwidth]{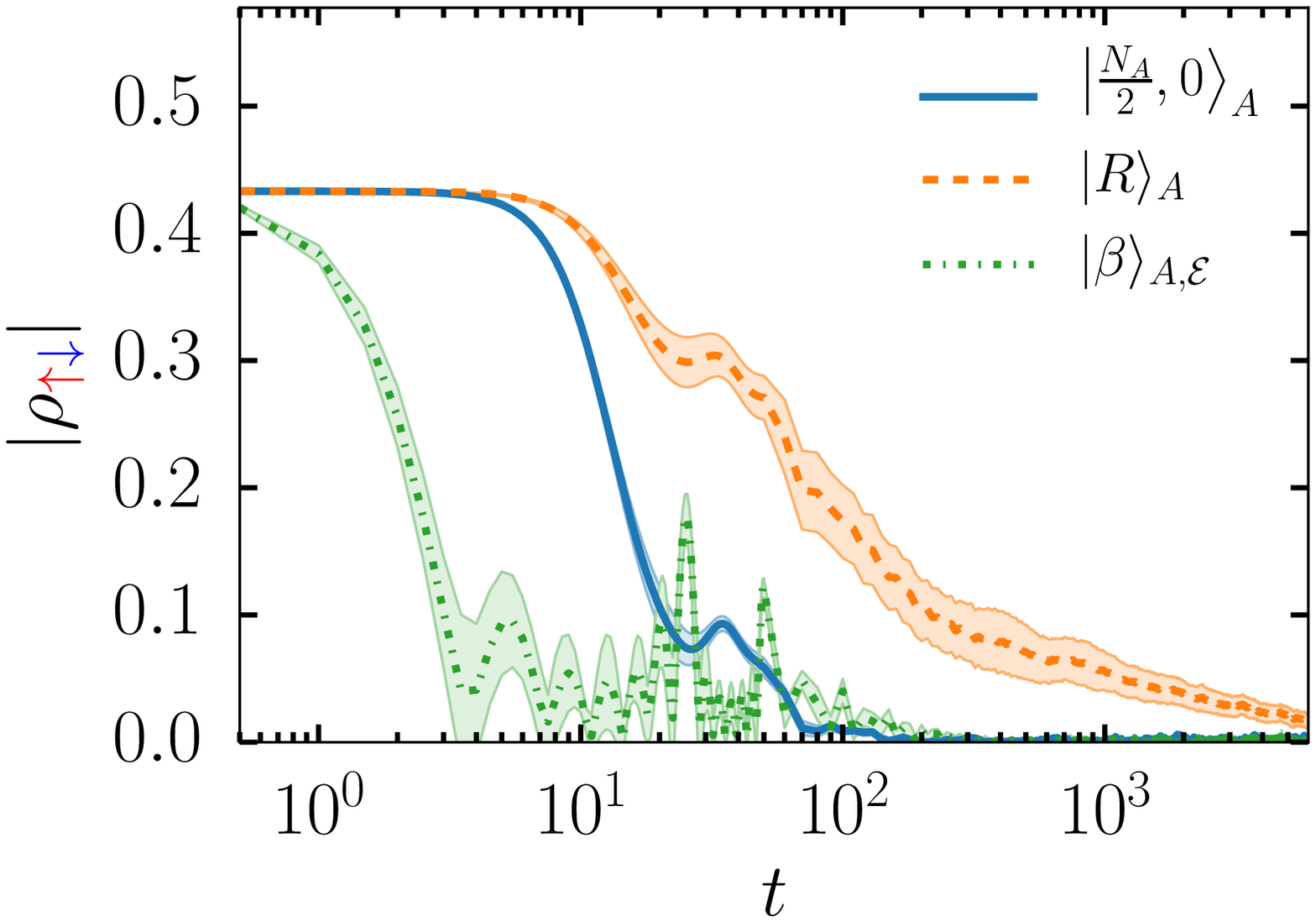}
		\end{center} 
		 & 
		 \vspace*{-1.3cm}
		 \begin{center}
			\includegraphics[width=0.39\textwidth]{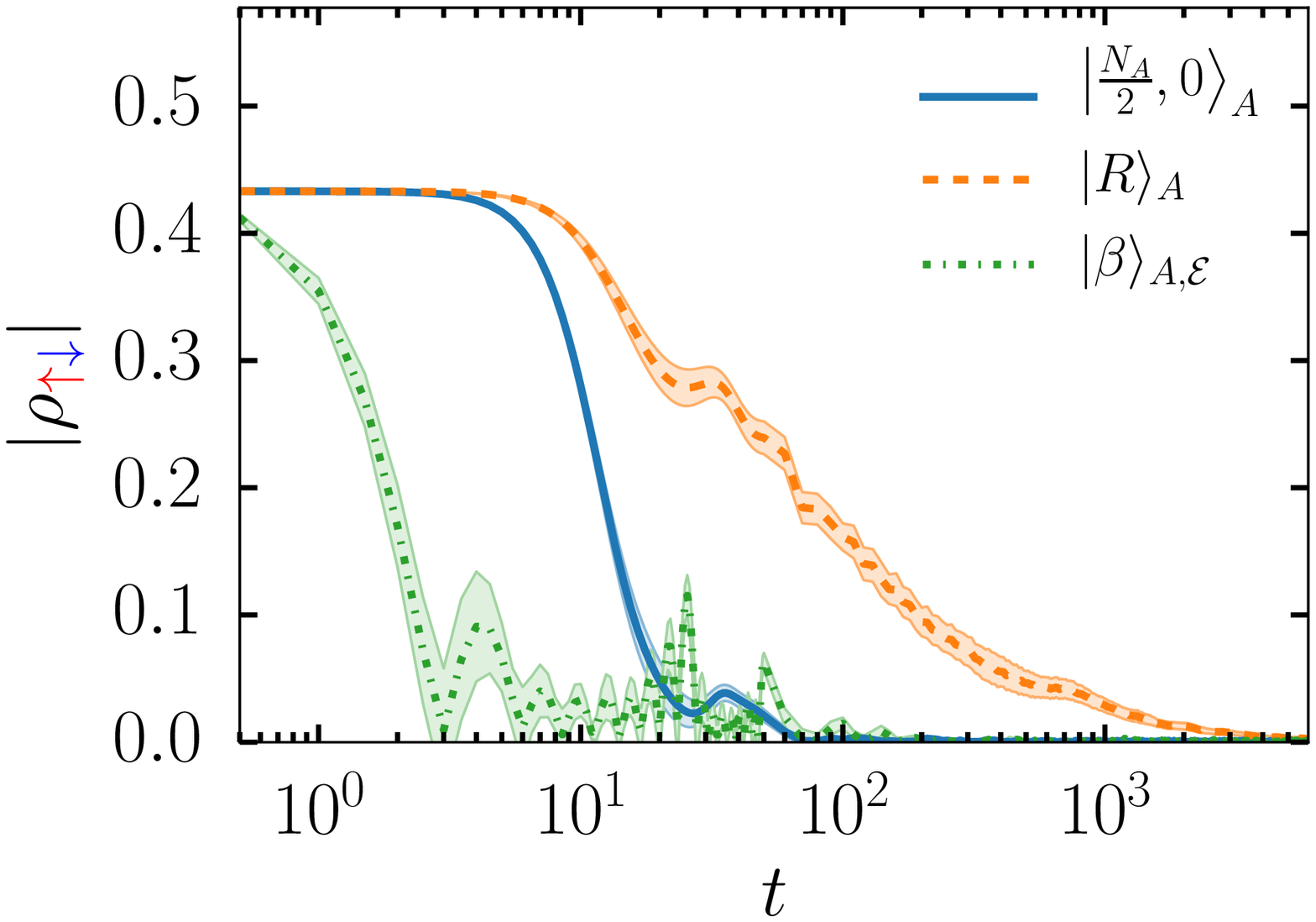}
		 \end{center}
		\\
	\end{tabular}
	\end{adjustbox}
	\caption{Development of system-apparatus correlation (top row) and the loss of phase coherence (bottom row) for test object $\left|\psi(\frac{3}{4})\right\rangle_S$. The size of the apparatus $N_A$ is indicated in the columns, and the initial apparatus-environment (ready) states are shown in the panels. The average (lines) and the region within one standard deviation (shading) are calculated by averaging over $n_r=15$ realisations of the ready state.}
	\label{fig:dynamics_apparatus}
\end{figure*}

In this section, the focus shall be on the ability of the measurement apparatus to quench the phase coherence of spin $S$ and the capacity to develop system-apparatus ($S$-$A$) correlations~\cite{PERE86}. To this end, the reduced density matrix (RDM) of the $S$-$A$ combination is introduced
\begin{equation}
\rho_{SA}(t) = \mathrm{Tr}_{\cal E} \left[ |\Psi(t)\rangle\langle \Psi(t)| \right] \, ,
\end{equation}
where the trace is over all spins in ${\cal E}$ ($N_{\cal E}$ in total).
The phase coherence corresponding to $S$ is then
\begin{equation}
\rho_{\uparrow \downarrow} = \rho_{\downarrow \uparrow}^* = \mathrm{Tr}_{A} \left[ {}_S\langle \uparrow \mid \rho_{SA} \mid \downarrow \rangle_S \right] \, ,
\end{equation}
where the trace is now over the $N_A$ apparatus spins.

The simulation results are shown in Fig.~\ref{fig:dynamics_apparatus}, where the data is averaged over $n_r=15$ realisations of the apparatus-environment ($A$-${\cal E}$) ready state. For convenience, the correlations are expressed in terms of Pauli matrices (i.e., $2\bm{S}_{S/A}= \bm{\sigma}_{S/A}$) and normalised with $N_A$.
Let us highlight some key observations: 
\begin{enumerate}
\item The apparatus initial state $|\frac{N_A}{2},0\rangle_A$ and the $A$-${\cal E}$ state $|\beta\rangle_{A,{\cal E}}$ lead to significant, but not maximal, correlation between the system and the apparatus (maximal correlation corresponds to $\langle \sigma^z_S \sigma^z_A\rangle/N_A = 1$). In comparison, the $S^z_A=0$ random state $|R\rangle_A$ develops only a modest amount of correlation. To be more precise: For $|\frac{N_A}{2},0\rangle_A$ the correlation saturates around $0.78$, $0.79$, and $0.76$ of its maximum value for respectively $N_A=4$, 6, and 8, for the data shown in the figure.
\item While the loss of coherence is an order of magnitude faster for $|\beta\rangle_{A,{\cal E}}$ compared to $|\frac{N_A}{2},0\rangle_A$, the development of correlations are rather similar. Presumably, the initial $A$-${\cal E}$ entanglement of $|\beta\rangle_{A,{\cal E}}$ enhances decoherence, but does not affect relaxation.
\item The decoherence time scale progressively decreases upon increasing the apparatus size $N_A$.
\end{enumerate}
Note further that the large standard deviation in the $S$-$A$ correlation for $|R\rangle_A$ and the coherence recurrences for $|\beta\rangle_{A,{\cal E}}$, as both observed for $N_A=4$, appear to be small apparatus-size artefacts; these characteristics are reduced upon increasing $N_A$. In contrast, the standard deviation for $\PLRZ$ is almost always smaller than the linewidth, for each $N_A$ shown.
Correlations $\langle \sigma^\alpha_S \sigma^\alpha_A\rangle$ along $\alpha=x,y$ originate from fluctuations in the apparatus (spin $S$ is a constant of motion), are essentially zero, and hence not shown here.

\begin{figure*}
\begin{adjustbox}{center}
	\begin{tabular}{m{0.03\textwidth}|*{3}{m{0.38\textwidth}}}
		& \centering{$N_A=4$} & \centering{$N_A=6$} & \hspace{2cm}$N_A=8$ \\
		 \hline 
		 \renewcommand{\arraystretch}{0.0}% Tighter
		\rotatebox{90}{Entropy} &  %\vspace*{0.cm}
		\begin{center}
			\includegraphics[width=0.39\textwidth]{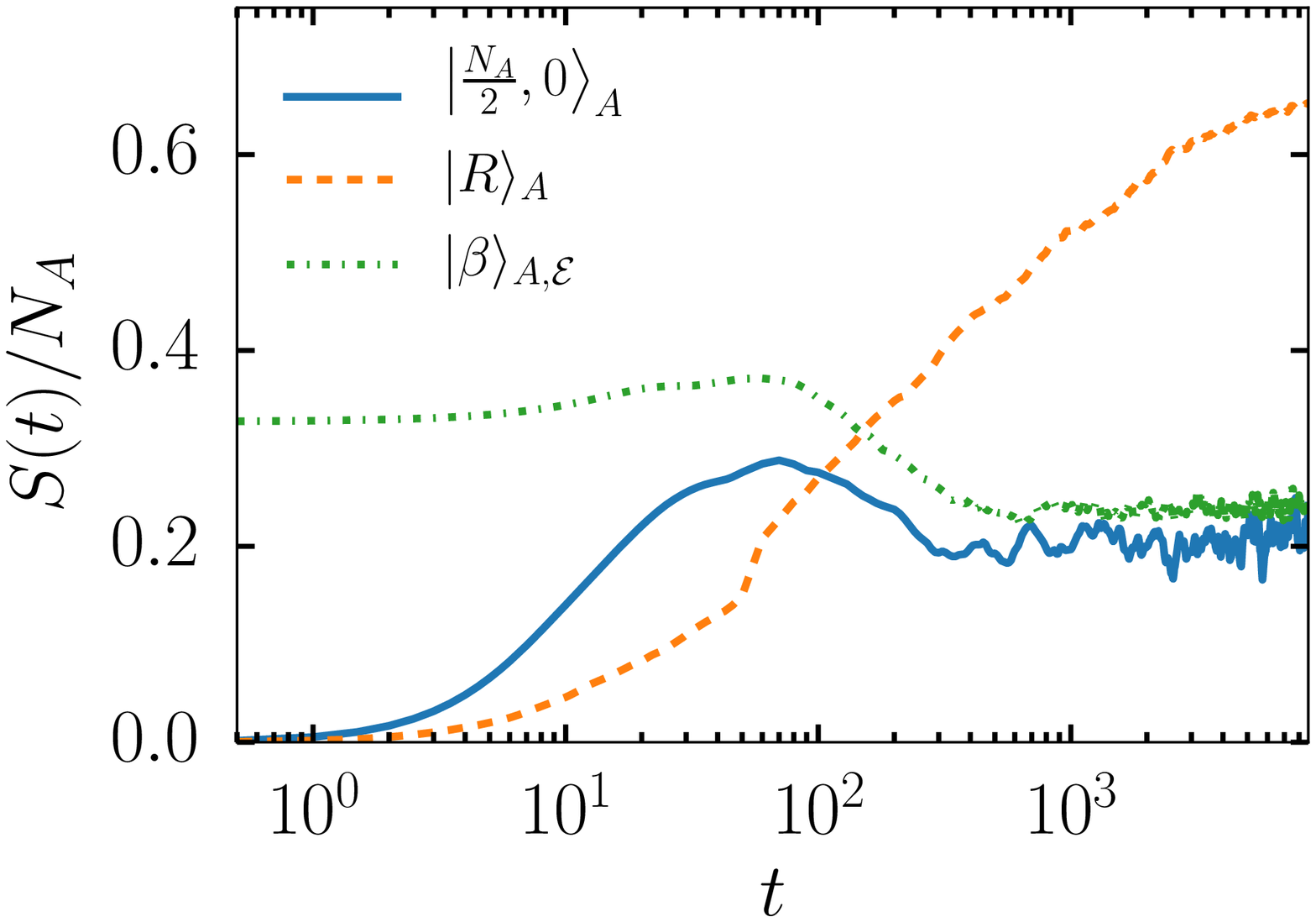}
		\end{center} 
		&
		\begin{center}
			\includegraphics[width=0.39\textwidth]{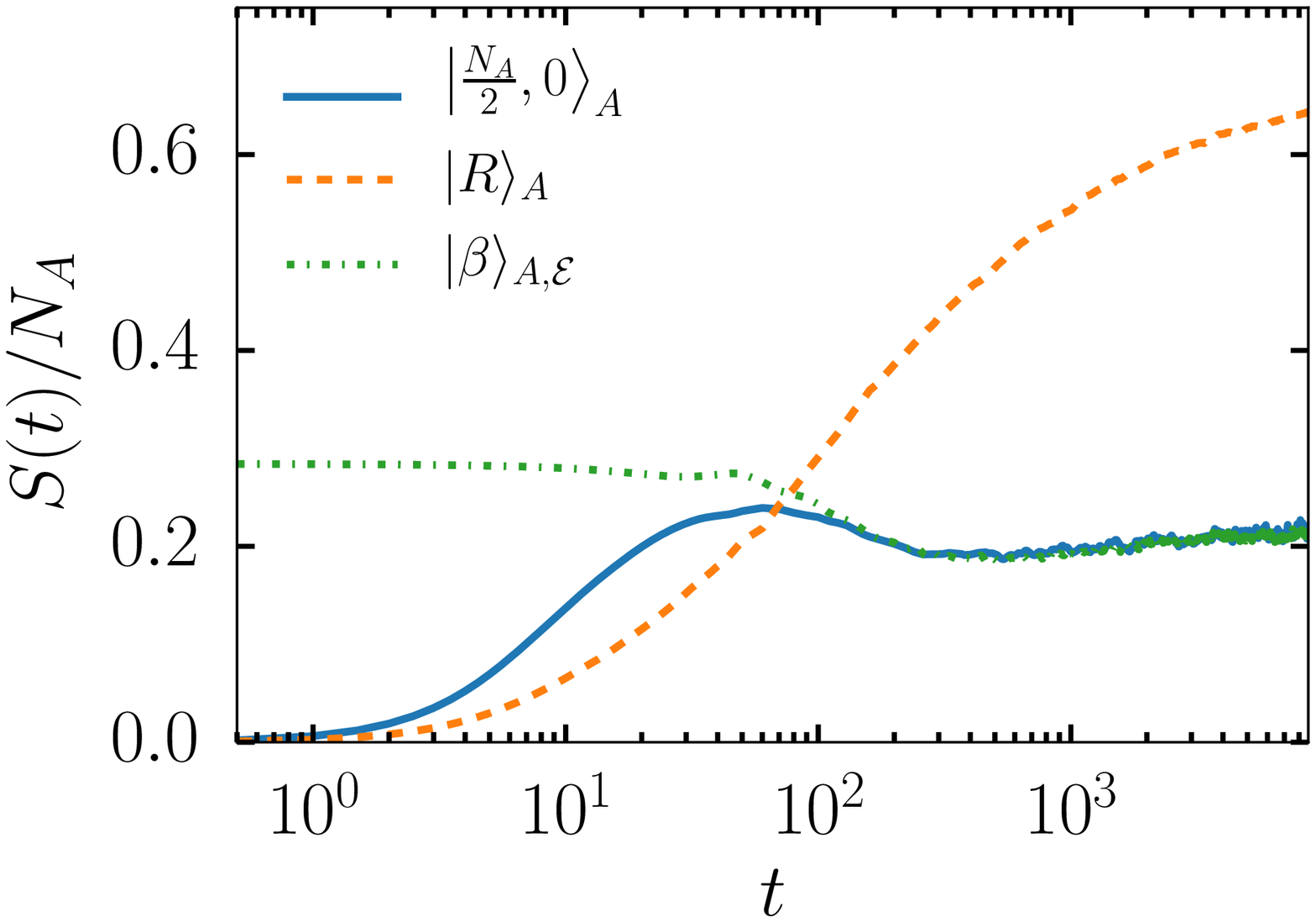}
		\end{center} 
		 &
		\begin{center}
			\includegraphics[width=0.39\textwidth]{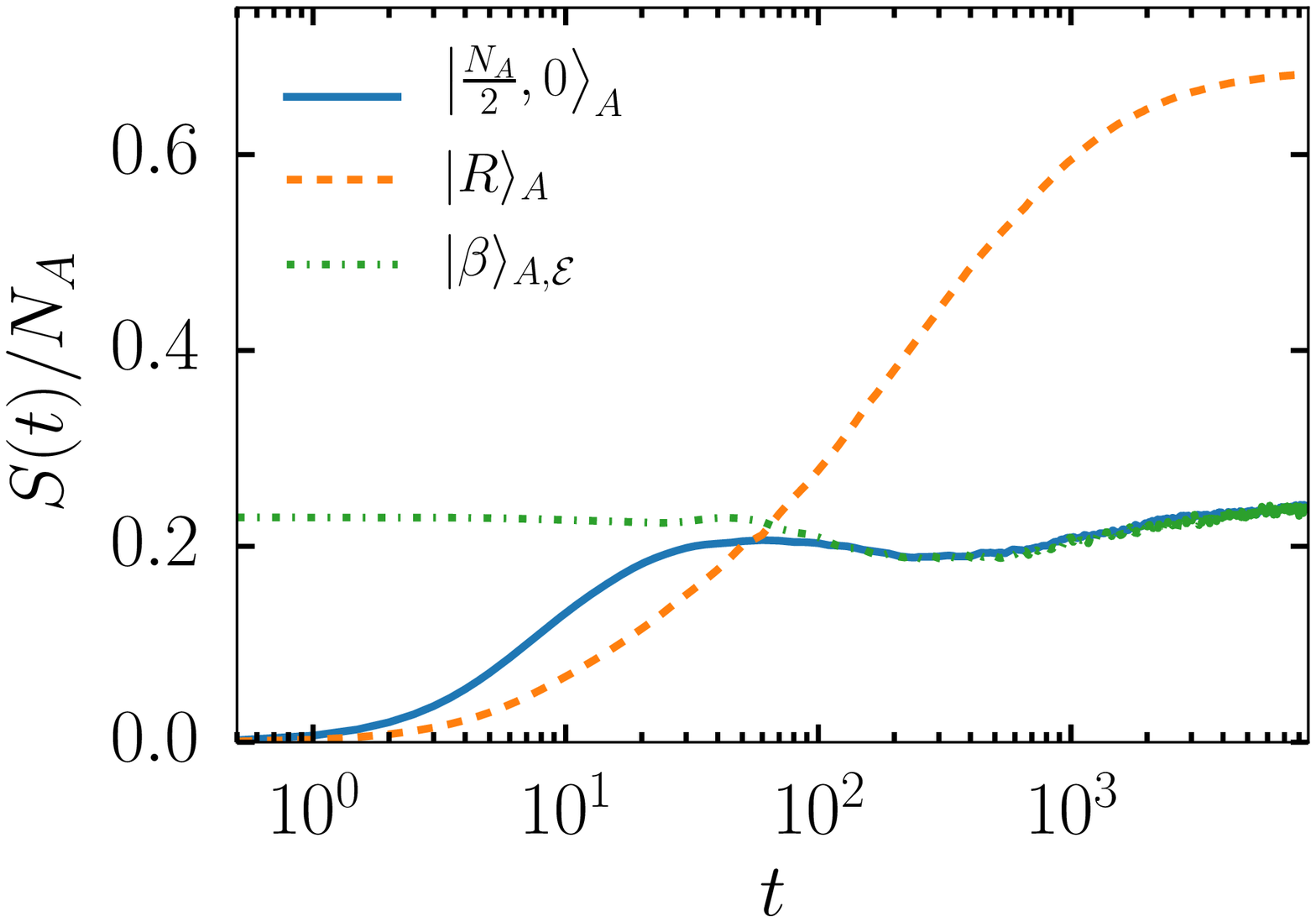}
		\end{center}
		\\
	\end{tabular}
\end{adjustbox}
	\caption{Entropy $S(t)$ of the apparatus relative to the spin-up state of the test object---calculated using the reduced density matrix $\tilde{\rho}_{\uparrow\uparrow}(t)$, see Eq.~(\ref{eq:rdm})---as a function of time.  The apparatus size $N_A$ is indicated above each column. 
	Initially, the test object is prepared in $\left|\psi(\frac{3}{4})\right\rangle_S$ with the remainder indicated in the legend.}
	\label{fig:entropy}
\end{figure*}
Since initial state $|\frac{N_A}{2},0\rangle_A$ does not saturate to its maximal value, the apparatus does not perform a simple precession in the maximal multiplicity subspace. But one might one wonder whether it is still possible to ascribe a pure state (the pointer reading) to the apparatus relative to the state of spin $S$. Therefore, 
the entropy of the RDM corresponding to state $i$ of S
\begin{equation}\label{eq:rdm}
	\rho_{ii} = {}_S\langle i|\rho_{SA}|i\rangle_S \, ; \quad \tilde{\rho}_{ii} = \rho_{ii} / \mathrm{Tr}[\rho_{ii}] \, ,
\end{equation}
is calculated, with $i$ equal to $\uparrow$ or $\downarrow$. The entropy corresponding to density operator $\rho(t)$ is defined as 
\begin{equation}\label{eq:entropy}
S(t) = -\mathrm{Tr}\left[ \rho(t) \ln \rho(t)\right] \, .
\end{equation}
Fig.~\ref{fig:entropy} shows, for each initial state, the entropy time development of a single (representative) simulation. 
Interestingly, the entropy of initial state $|\frac{N_A}{2},0\rangle_A$ thermalises to that of initial state $|\beta\rangle_{A,{\cal E}}$ after some initial relaxation---the non-zero entropy $S(t=0)$ of $|\beta\rangle_{A,{\cal E}}$ originates from the initial $A$-${\cal E}$ entanglement of that state. 
As expected for high-temperature ready state $|R\rangle_A$, the apparatus' entropy turns out significantly more than the other two states (almost a factor 3 in the data displayed). 
Observe, moreover, the entropy decrease for $|\beta\rangle_{A,{\cal E}}$, especially near $t \approx 10^2$, which is indicative for non-Markovian behaviour~\cite{BREU07}.
In line with the apparatus being extensive, the entropy scales with $N_A$ for all three ready states.
The data for RDM $\tilde{\rho}_{\downarrow\downarrow}(t)$ does not differ in an essential way from Fig.~\ref{fig:entropy}.

Data not shown here indicates that the coherence of $\tilde{\rho}_{ii}(t)$, in the basis diagonalising $H_A$, is only partly reduced. The remaining coherence is amongst the states with maximal multiplicity $S_A = N_A/2$, which are degenerate in $H_A$. In particular, coherence in $\tilde{\rho}_{\uparrow\uparrow}$ ($\tilde{\rho}_{\downarrow\downarrow}$) is between those states with $S_A = N_A/2$ and near-maximum (near-minimum) magnetisation. As a result, a purely statistical description using a partition function is inadequate to capture all facets of our apparatus.

Since $|R\rangle_A$ poorly correlates with the apparatus and displays suboptimal truncation of spin $S$, the focus shall be on $\PLRZ$ and $\BETA$ in the remainder of the text.

\section{Calibration}\label{sec:calibration}
\begin{figure}
	\centering
	\subfloat[]{
	\includegraphics[scale=0.25]{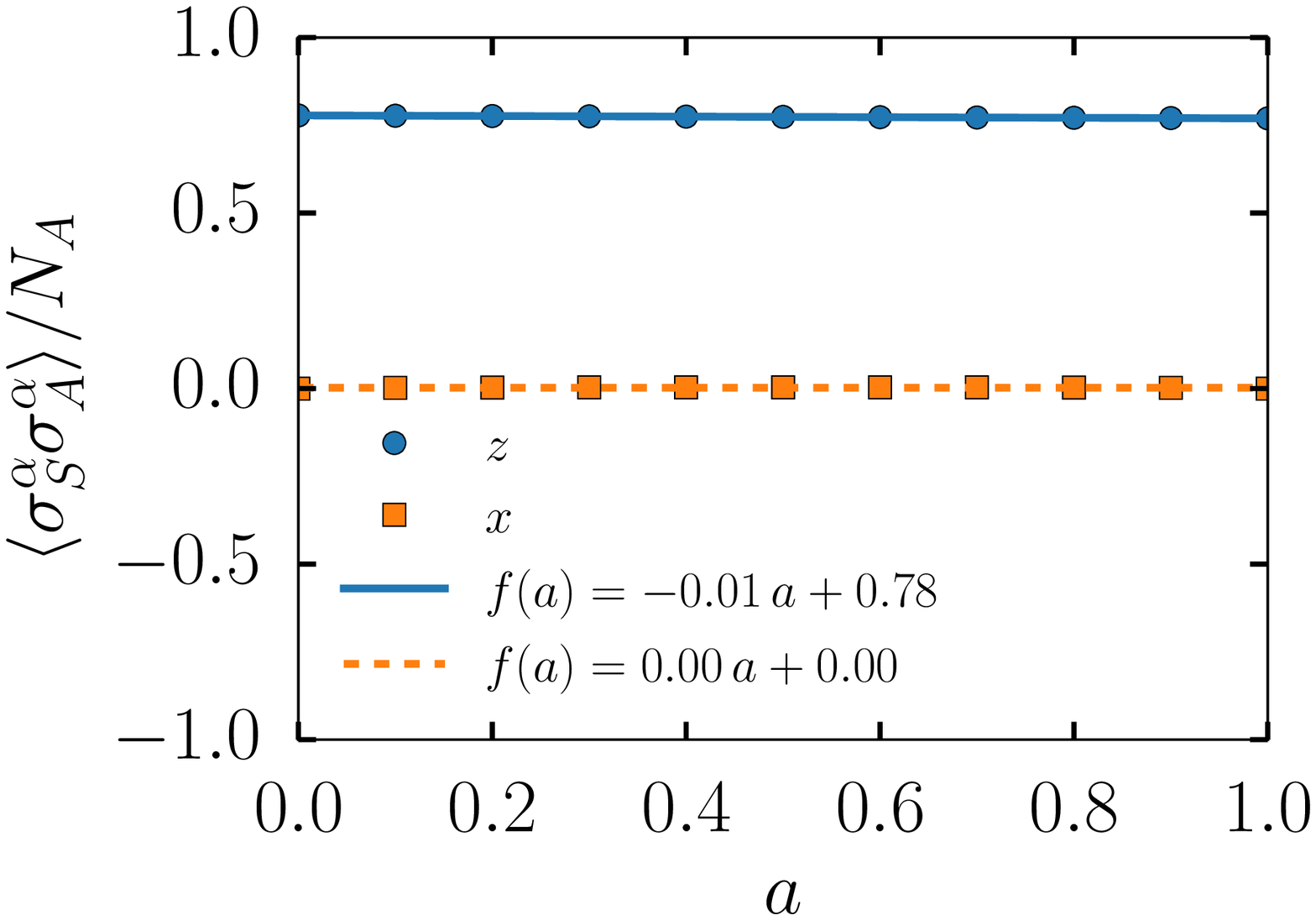}}
	\qquad
	\subfloat[]{
	\includegraphics[scale=0.25]{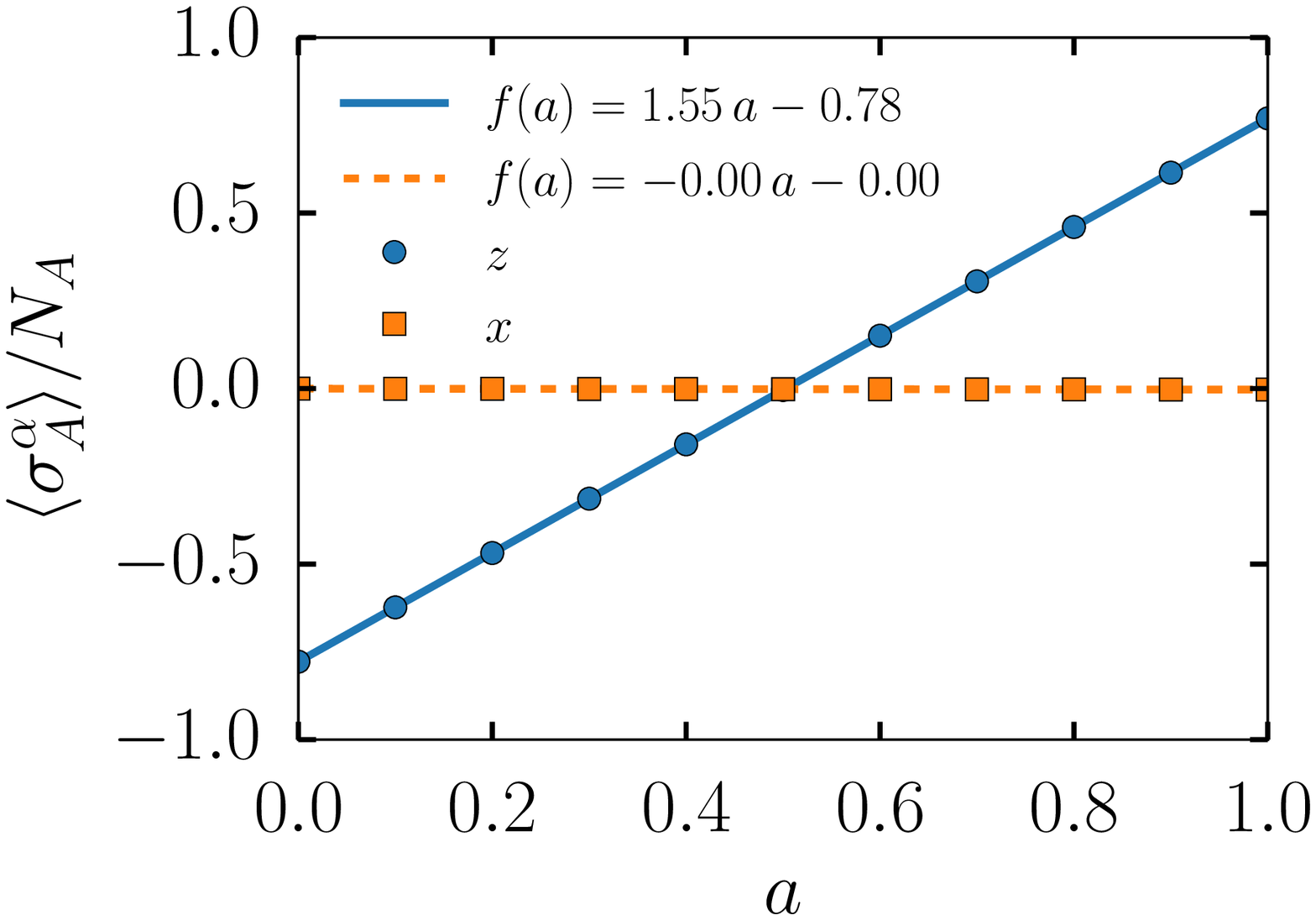}
	}
	\caption{System-apparatus correlation $\langle \sigma_S^\alpha \sigma_A^\alpha \rangle$ (left) and the apparatus magnetisation $\langle \sigma_A^\alpha \rangle$ (right) with $\alpha$ indicated in the panels. 
	Markers indicate simulation data for initial state $|\Psi(0)\rangle = |\psi(a)\rangle_S \left|\frac{N_A}{2},0 \right\rangle_A |\beta\rangle_{\cal E}$, after the apparatus (of $N_{A}=4$ spins) has relaxed at $t=10^4$. The lines are fits to the data points.}
	\label{fig:a_sweep}
\end{figure}
Almost every pair of quantum states that are led to interact, will produce a mutual imprint. This imprint is in general quite intricate and non-generic, which makes it---in practice---difficult to use the imprint to infer the original states before interaction took place. 
Hence, a measurement device is not only required to measure spin $S$, but do so in a way that allows one to infer the state of the to-be measured object.
The apparatus will now be examined to see if it gives an unbiased indication of the value $a$ in Eq.~(\ref{eq:S_a}).

In Fig.~\ref{fig:a_sweep}, the $S$-$A$ correlation (left panel) and order parameter (right panel) are plotted against spin $S$ parameter $a$ with initial apparatus state $\left|\frac{N_A}{2},0 \right\rangle_A$ and $N_A=4$.
The data indicates that i) the apparatus does not have a bias towards a particular value of $a$, as shown in the left panel and ii) the order parameter $\langle \sigma_A^z\rangle$ (right panel) can directly be used to infer $a$.

On average, the same unbiased behaviour is observed for initial state $|\beta\rangle_{A,{\cal E}}$ with an average standard deviation $\mathrm{std}[\langle \sigma_S^z \sigma_A^z \rangle]$ of 0.08 (averaged over $a$) for $n_r=30$ realisations per $a$ and (to preserve computation time) $N_{\cal E}=11$ instead of 12. Individual runs, however, \emph{can} show a bias towards low or high $a$ as a result of initial non-zero apparatus magnetisation in $|\beta\rangle_{A,{\cal E}}$. But these artefacts are expected to disappear upon increasing $N_A$.
 
Finally, it was observed that, in line with the data in Sec.~\ref{sec:truncation}, 
the size of the correlations scale linearly with $N_A$.
The ability to unbiasedly capture the correlation and order parameter with a linear fit directly carries over to $N_A=6$ and 8, for both $\PLRZ$ and $\BETA$.

\section{Stability test}\label{sec:stability}
\begin{figure}
	\centering
	\includegraphics[scale=0.3]{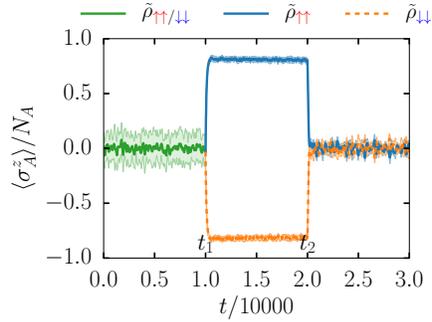}
	\caption{Time evolution of the order parameter $\mathrm{Tr}\left[ \sigma^z_A \tilde{\rho}_{ii} \right]/N_A$ of the apparatus relative to spin $S$ state $i$ [see also Eq.~(\ref{eq:rdm})]. The apparatus ($N_A=4$ spins) is connected to spin $S$ (initial state $|\Psi(0)\rangle = \left|\psi\left(\frac{3}{4}\right)\right\rangle_S \BETA$) at $t_1=10^4$ until $t_2=2\cdot 10^4$ [see Eq.~(\ref{eq:ISA_quench})]. The average (lines) and the region within one standard deviation (shading) are calculated by averaging over $n_r=15$ realisations of the ready state $\BETA$.}
	\label{fig:N4_quench}
\end{figure}

\begin{figure}
	\centering
	\includegraphics[scale=0.3]{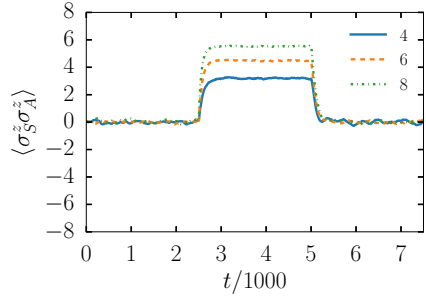}
	\caption{Time evolution of the system-apparatus $z$-correlation for initial state 
	$|\Psi(0)\rangle = \left|\psi\left(\frac{3}{4}\right)\right\rangle_S \left|\frac{N_A}{2},0 \right\rangle_A |\beta\rangle_{\cal E}$ with $N_A$ indicated in the legend. Spin $S$ is connected at $t_1= 2.5\cdot 10^3$ to the apparatus and decoupled at $t_2=5 \cdot 10^3$.}
	\label{fig:quench_correlation}
\end{figure}

The goal of this section, to evaluate the stability of the initial states, is two-fold.
Firstly, the free evolution of an initial state is examined to explore whether this spoils its ability to give pointer readings and if it leads to false positives.
Secondly, the stability of the apparatus readings are analysed, to see if it is able to retain its registration record upon completion of the measurement. 
To this end, spin $S$ is coupled to the apparatus in the time window $t_1 \leq t \leq t_2$ only. Thus, the $S$-$A$ coupling $I_{SA}$ is made time dependent as
\begin{equation}\label{eq:ISA_quench}
I_{SA}(t) = \left\{ \begin{matrix} I_{SA} & t_1 \leq t \leq t_2\\
0 & \mathrm{otherwise} \end{matrix} \right. \, .
\end{equation}
In Fig.~\ref{fig:N4_quench} the order parameter is shown relative to its spin $S$ state, with initial state $\BETA$. No distinction can be made between $\tilde{\rho}_{\uparrow\uparrow}$ and $\tilde{\rho}_{\downarrow\downarrow}$ before connecting at $t < t_1$, since the system and apparatus still form a product state. 
Large fluctuations in the order parameter are observable as indicated by the green shading. Nevertheless, after $t_1$ the development of magnetic order relative to the spin $S$ state is unaffected by the initial evolution. While this is perhaps expected for the state $\BETA$ since it is supposed to resemble a thermal (and therefore steady-) state of the apparatus and environment combined, the same holds for $\PLRZ$. Fig.~\ref{fig:quench_correlation} depicts the $S$-$A$ correlation, in which the same behaviour is observed for $\PLRZ$ (but now with $t_1= 2.5\cdot 10^3$ and $t_2=5 \cdot 10^3$). No false positive measurements were observed in any of the simulations.

Finally, and perhaps most unexpectedly, upon decoupling the apparatus at $t_2$ the correlations are immediately washed out (see both Figs.~\ref{fig:N4_quench} and \ref{fig:quench_correlation}). The apparatus is thus not able to accommodate stable pointer readings upon completion of the measurement. In Ref.~\cite{ALLA13} similar observations, of ineffective registration, were reported for i) a two spin ferromagnet unless the temperature is very low and ii) a macroscopic ferromagnet undergoing a first-order phase transition, where the pointer readings become stuck in a paramagnetic fixed point (and return to zero magnetisation upon decoupling of the test spin) if the system-apparatus coupling is too weak. 

While one might expect the states of our apparatus to stabilise upon increasing $N_A$, our calculations show that it is not the case for $N_A=6$ and $N_A=8$. This is true for both $\BETA$ and $\PLRZ$, with the results of the latter shown in Fig.~\ref{fig:quench_correlation}. Introducing a near-neighbour anisotropy to the apparatus $H_A^\prime = H_A - \Delta \sum_{i=2}^{N_A}S_i^z S_{i+1}^z$ to help pin the fully polarised states, does not stabilise the pointer readings upto $N_A=8$. Instead it suppresses the ability to develop system-apparatus correlations in the first place.
Neither is replacing the near-neighbour chain $H_A$ by a fully connected magnet $\tilde{H}_A = -J/N  \bm{S}_A \cdot \bm{S}_A$. The same instability is found, which can be understood from the fact that $\tilde{H}_A$ has the same $N_A+1$-fold degenerate ground state subspace as $H_A$. Here too the introduction of anisotropic terms (for each connection) does not salvage the ferromagnetic configuration, and the same suppression of $S$-$A$ correlations is found as for $H_A$.

\section{Discussion and Conclusion}\label{sec:disc_concl}
In this work, a measuring instrument was constructed that aims to measure the magnetic moment of $S$, a spin-1/2 particle. The goal was to build a device that exploits the sensitivity of an initial state that is susceptible to symmetry breaking to, in this way, amplify a microscopic signal. The device consists of a ferromagnetic chain $A$ that is immersed in a low-temperature thermal reservoir ${\cal E}$. By coupling the $z$-component of spin $S$ to the order parameter of $A$, the test object leads to explicit symmetry breaking of the ferromagnet.
In contrast to earlier work~\cite{ZIMA88,GAVE90,FIOR94,ALLA03, ALLA13}, the present results account for the full quantum many-body dynamics without resorting to mean field and/or quasi-classical approximations. The turn side is that the instrument is not really macroscopic and contained up to $N_A+N_{\cal E}=20$ spins only.

It was found that the device can develop pointer readings with significant, but not maximal, correlation to $S$. In the process, the coherence of the reduced density matrix of $S$ is quenched, thereby leading to a mixed state. Furthermore, the state of the ferromagnet relative to either spin up or down of $S$ is itself not pure but, instead, described by a mixed state.

Going further, the instrument was found to give unbiased measurements of $S$ and no false positive readings were observed. Finally, the simulations indicated that the apparatus was unable to irreversibly register the measurement outcomes. Meaning that, upon finishing the measurement, the device was unable to maintain its record. It is expected that this characteristic is a peculiarity of the small size of the ferromagnet $N_A$, but a comparison of $N_A=4$, 6, and 8 are unable to substantiate this view. 

With the danger of stating the obvious, we note that this work has no bearing on the quantum measurement problem---i.e., the occurrence of individual events---for this would require additional interpretative elements~\cite{ALLA13}. At present, these interpretational elements are not amendable to objective and independent verification and are, as such, beyond the scope of this work.

\section{Acknowledgements}
MIK and HCD acknowledge financial support by the European Research Council, project
338957 FEMTO/NANO. We are grateful to H. J. Kappen for useful suggestions.

\appendix
\section{Fine tuning the coupling strengths}\label{sec:tuning}
\begin{figure*}
\begin{adjustbox}{center}
	\subfloat[System-apparatus coupling]{
		\includegraphics[scale=0.27]{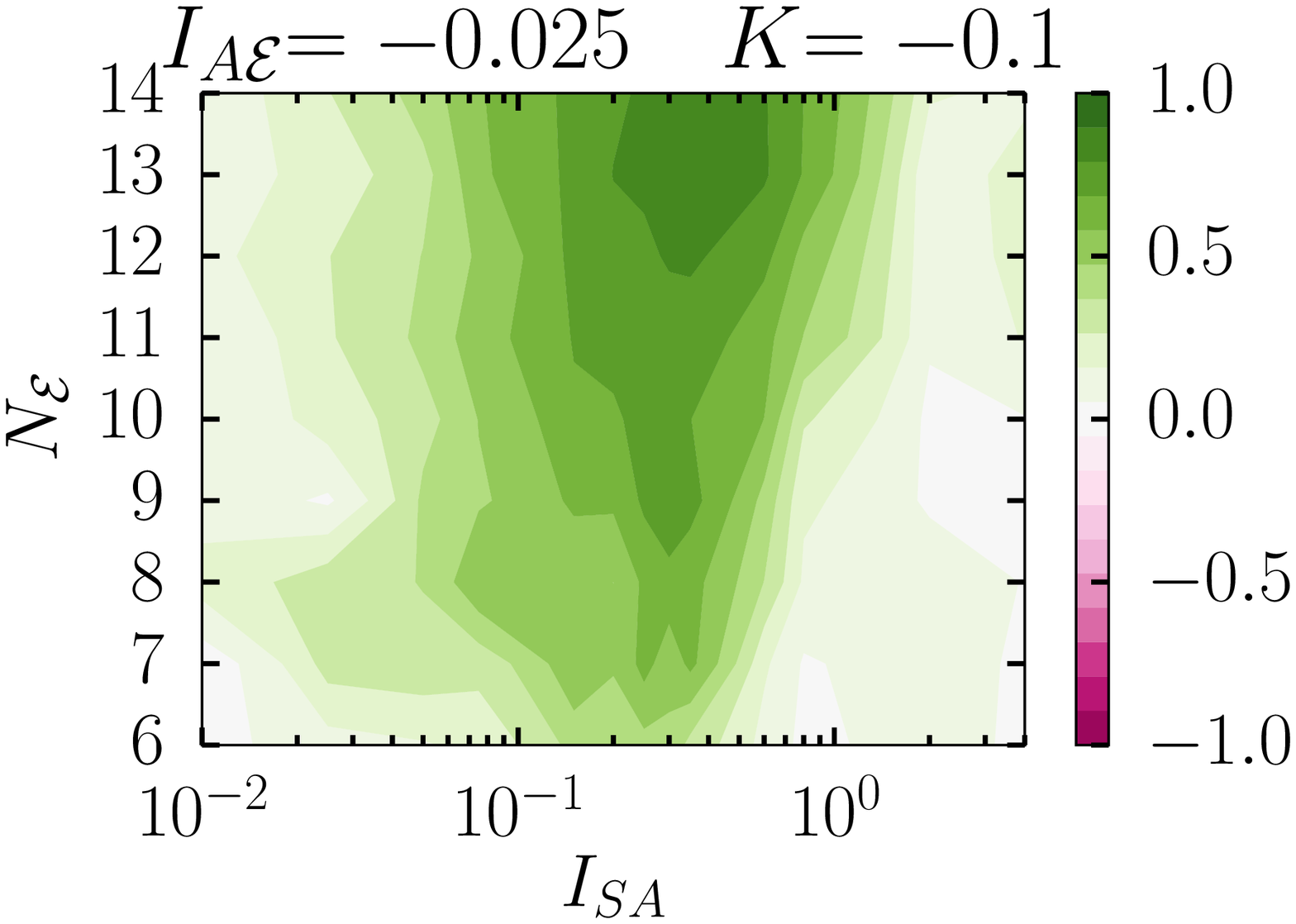}
	}
	\subfloat[Apparatus-environment coupling]{
		\includegraphics[scale=0.27]{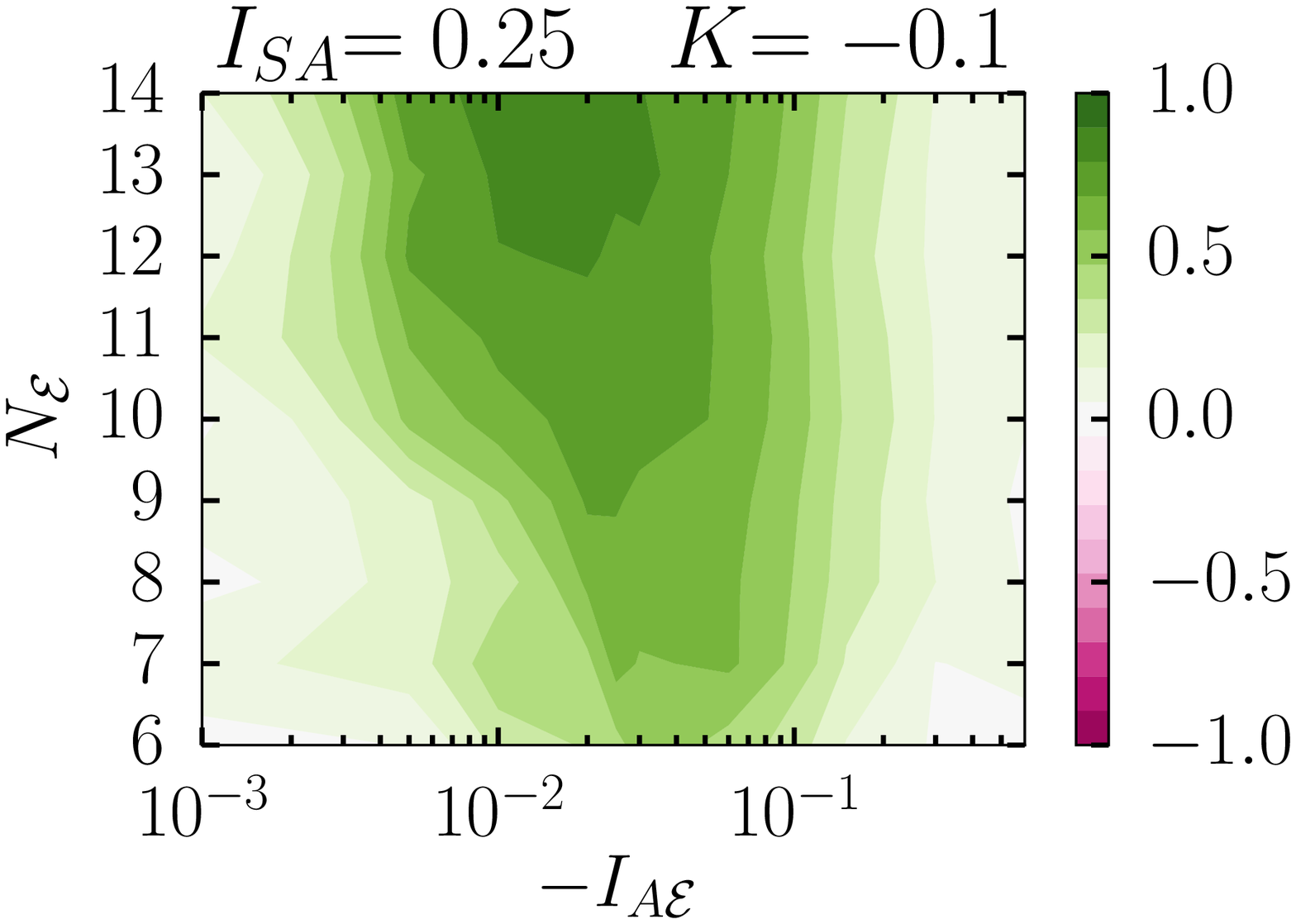}
	}
	\subfloat[Intra-environment coupling]{
	\includegraphics[scale=0.27]{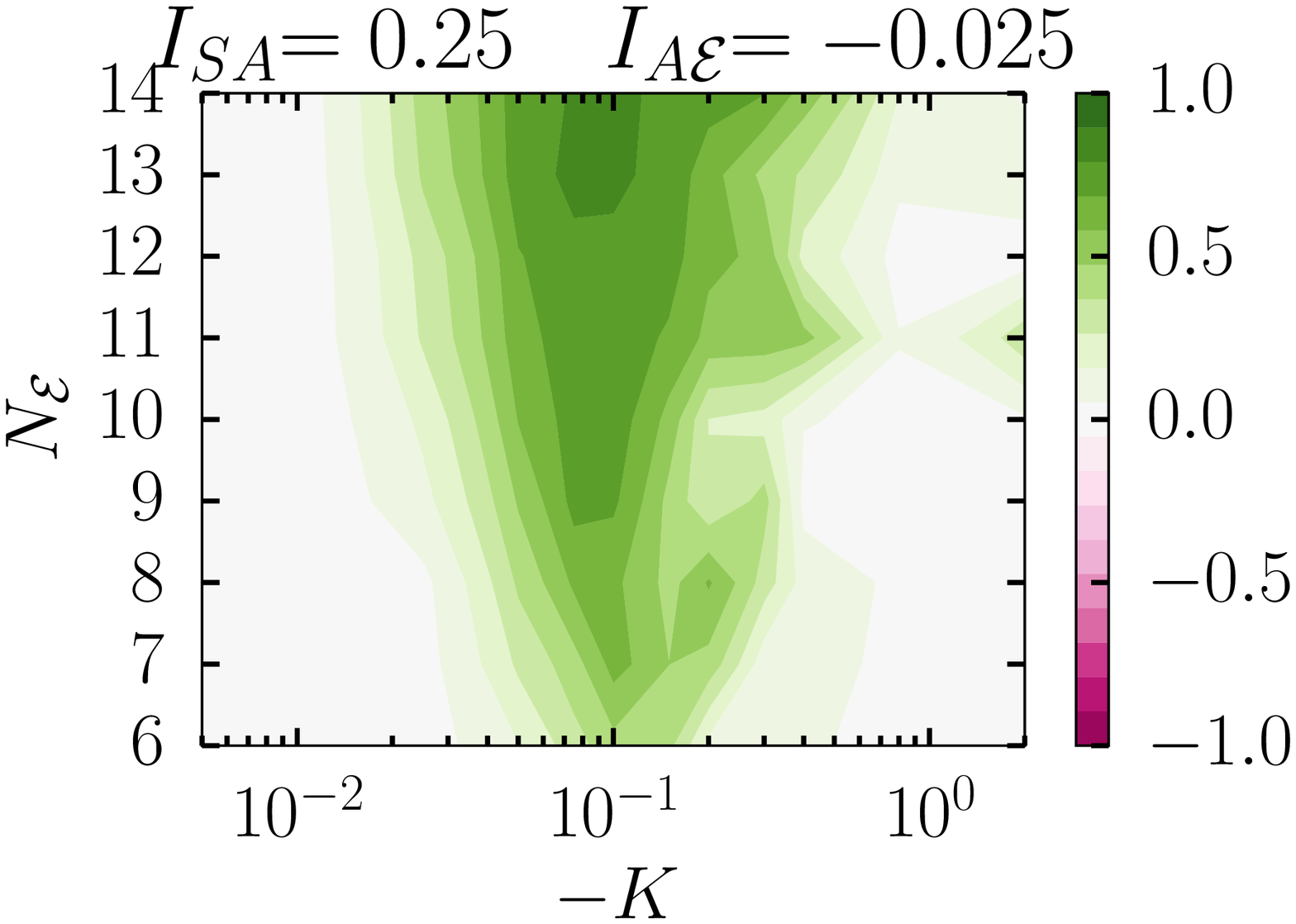}
	}
\end{adjustbox}
	\caption{Contour plots of the system-apparatus correlation $\langle S_S^z S_A^z\rangle$ evaluated at time-step $t=10^3$ with $N_A=4$. The panels indicate that for a given environment size $N_{\cal E}$ there is a narrow region in the coupling strength (expressed in units of $J$ and shown in logarithmic scale) that lead to appreciable system-apparatus correlations. The region expands upon enlarging the size of the environment.}
	\label{fig:parameter_window}
\end{figure*}

For efficient relaxation in small spin systems it is not only beneficial to use spin glass types of environments~\cite{YUAN07}, but also to use finely tuned couplings strengths. For spin environments of finite size, there is a small window in the parameters that lead to optimal relaxation, as already pointed out in several other works ~\cite{YUAN07, JIN10, DONK18}. In general, one might hope that the loss in generality of the specific coupling strengths is an artefact of the small size of the environment. Meaning, the constraints on the values of the parameters disappear when the environment size becomes sufficiently large. Thus, by doing small spin simulations one sacrifices the generality in the coupling strengths. Extensive number of simulations for the measurement set up [see Eq.~(\ref{eq:H}) for the Hamiltonian] whereby all free parameters---to wit, the system-apparatus coupling $I_{SA}$, the apparatus-environment coupling $I_{A{\cal E}}$, and the intra-environment coupling $K$---were varied corroborate the hypothesis that the window of optimal parameter values expands upon increasing the environment size $N_{\cal E}$. The results are shown in Fig.~\ref{fig:parameter_window} whereby the system-apparatus correlation, $\langle S_S^z S_A^z\rangle$, is shown for 3 slices of parameter space of the set $\{I_{SA}, I_{A{\cal E}}, K \}$ with initial state $\PLRZ$ and $N_A=4$. 
Additional calculations (not shown here) for $\BETA$ with $N_A=4$ and $N_A=6$ indicate qualitatively the same, but slightly less sharp, sensitivity for the couplings compared to $\PLRZ$.

%\section*{References}
\bibliographystyle{elsarticle-num}
\bibliography{refs}
\end{document}